\documentclass{article}
\pdfoutput=1 

\usepackage{graphicx}
\usepackage{amssymb, amsmath, amsthm, amsfonts, bm, mathrsfs, wasysym, txfonts, bbm, enumerate}

\begin{document}

\title{Quantum vorticity in nature\thanks{%
Invited contribution to \textit{The} \textit{Proceedings of a Conference on
Sixty Years of Yang-Mills Gauge field}, Institute of Advanced Studies,
Nanyang Technological University, Singapore, May 26-28, 2015.}}
\author{Kerson Huang \\
Physics Department\\
Massachusetts Institute of Technology\\
Cambridge, MA 02139 USA}
\maketitle

\begin{abstract}
Quantum vorticity occurs in superfluidity, which arises from a spatial
variation of the quantum phase. As such, it can occur in diverse systems
over a wide range of scales, from the electroweak sector and QCD of the
standard model of particle theory, through the everyday world, to the
cosmos. I review the observable manifestations, and their unified
description in terms of an order parameter that is a complex scalar field.
\end{abstract}

\section{Overview}

Quantum vorticity is a manifestation of superfluidity, which arises from a
spatial variation of the quantum phase. This can occur in different types of
physical systems, but a unified description can be given in terms of an
order parameter that is a complex scalar field: 
\begin{equation}
\phi \left( x\right) =F\left( x\right) e^{i\sigma \left( x\right) }
\end{equation}%
Its existence signals the spontaneous breaking of a $U(1)$ gauge symmetry.

In atomic systems with Bose-Einstein condensation \cite{Dafolvo99}, the
broken gauge symmetry is global. In this case, $\phi $ represents the
condensate wave function, which can be described by the Gross-Pitaevskii
equation, a nonlinear Schr\"{o}dinger equation (NLSE):%
\begin{equation}
-\frac{\hbar ^{2}}{2m}\nabla ^{2}\phi +\left( g\phi ^{\ast }\phi -\mu
_{0}\right) \phi =i\hbar \frac{\partial \phi }{\partial t}  \label{NLSE}
\end{equation}%
where $m$ is the mass of the atoms, $g$ an interaction parameter, and $\mu
_{0}$ is the chemical potential.

In superconductivity \cite{DeGennes99}, the broken gauge symmetry is local,
and $\phi $ obeys a nonlinear Schr\"{o}dinger equation with gauge coupling: 
\begin{equation}
-\frac{\hbar ^{2}}{2m}\left( \nabla -\frac{i\hbar q}{c}\mathbf{A}\right)
^{2}\phi +\left( g\phi ^{\ast }\phi -\mu _{0}\right) \phi =i\hbar \frac{%
\partial \phi }{\partial t}  \label{GL}
\end{equation}%
where $q=2e$ is the charge of Cooper pairs, and $\mathbf{A}$ is a magnetic
field. This is called the Ginsburg-Landau (GL) equation.

In relativistic systems \cite{Xiong14}, these equations generalize to a
nonlinear Klein-Gordon (NLKG) equation :%
\begin{equation}
\square \phi +V^{\prime }\phi =J  \label{NLKG}
\end{equation}%
where $\square $ is the d'Alembertian in curved spacetime, which may contain
gauge couplings, and $J$ denotes a possible external current. The
self-interaction potential may be chosen to be a phenomenological $\phi ^{4}$
potential:%
\begin{equation}
V=\frac{\lambda }{2}\left( \phi ^{\ast }\phi -F_{0}\right) ^{2}+V_{0}
\end{equation}%
Thus 
\begin{equation}
V^{\prime }=\lambda \left( \phi ^{\ast }\phi -F_{0}\right)
\end{equation}%
where a prime denotes differentiation with respect to $\phi ^{\ast }\phi $.
Physically, $\phi $ may correspond to a component of the Higgs field in the
standard model, the order parameter associated with chiral symmetry breaking
in QCD, or Higgs-like fields in grand-unified or supersymmetric theories.

All these systems exhibit superfluidity, with superfluid velocity given by 
\cite{Huang14}%
\begin{equation}
\mathbf{v}_{s}=\kappa \nabla \sigma
\end{equation}%
where $\kappa =-c\left( \partial \sigma /\partial t\right) ^{-1}$ . In the
relativistic case, the requirement $|\mathbf{v}_{s}|$ $<c$ is guaranteed,
when $\partial ^{\mu }\sigma $ is time-like, i.e. ,$\partial ^{\mu }\sigma
\partial _{\mu }\sigma <0.$ In the nonrelativistic limit, $\sigma
\rightarrow -\left( mc^{2}/\hbar \right) t+\nu ,$and $\kappa \rightarrow
\hbar /m.$

The equations of motions can be rewritten in terms of $F$ and $\sigma $, and
the equation for $\sigma $ yields a equation for $v_{s}$ similar to the
classical Euler equation of hydrodynamics, with quantum corrections.
However, for many purposes it is simpler to stay with the original forms,
particularly numerical analysis.

Quantum vorticity arises from a quantization of the circulation, by virtue
of the continuity of $\phi $:

\begin{equation}
\oint\limits_{C}dx\cdot \nabla \sigma =2\pi n\text{ \ \ \ }\left( n=0,\pm
1,\pm 2,\ldots \right)
\end{equation}%
where $C$ is a closed loop. If $n\neq 0$, the loop encircles a vortex line
on which $\phi =0$. As illustrated in in Fig.1, the field rises from zero at
the line to an asymptotic value, with a characteristic healing length $\xi $%
. The vortex line can thus be thought of as a tube of radius $\xi $ in which
symmetry is restored, with expulsion of of the order parameter. The
superfluid velocity $v_{s}$ drops off with distance $r$ from the line center
like $r^{-1}$. Since the energy density of the superfluid flow is
proportional to $v_{s}^{2}$, the energy per unit length of of a single
straight vortex line diverges with the radius of the container. The
existence of vorticity depends on a vacuum field, hence on the nonlinearity;
it is absent in the usual Schr\"{o}diger equation, or Klein-Gordon equation,
where the field goes to zero at infinity.

\begin{figure}
[ptb]
\begin{center}
\includegraphics[
height=1.8in,
width=3.4765in
]%
{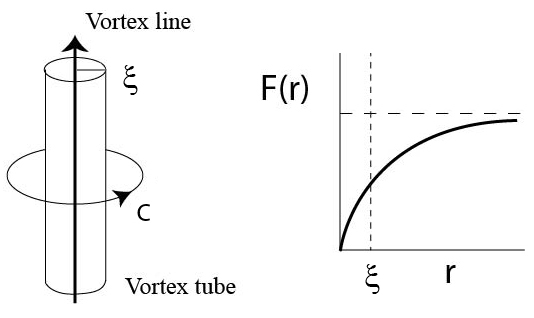}%
\caption{The vortex tube. The superfluid velocity decreases inversely with
distance from the central vortex line.}%
\end{center}
\end{figure}

A vortex line cannot terminate, except on boundaries, or upon itself,
forming a closed curve. The simplest example of the latter is vortex ring of
radius $R$, as illustrated in Fig.2. The ring moves with a translational
velocity normal to the plane of the ring$\ v\sim R^{-1}\ln R$ , with energy $%
E\sim R\ln R$. Thus, roughly speaking, $E\sim v^{-1}.$A vortex line of
arbitrary shape will move in such a manner that, at any point of the line,
the local velocity is normal to the tangent circle, with velocity inversely
proportional to the local radius of curvature. Thus, the line will generally
execute a self-driven writhing motion, as illustrated in Fig.2.

\begin{figure}
[ptb]
\begin{center}
\includegraphics[
height=1.8in,
width=3.4765in
]%
{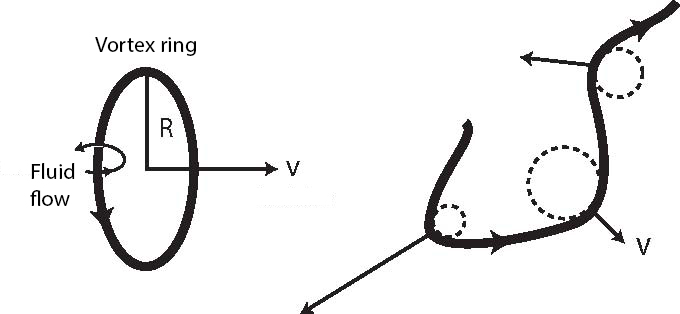}%
\caption{Left panel: The vortex ring has a
translational velocity $v$ approximately inversely proportional to its
radius $R$. Right panel: A general vortex line has local velocity normal to
the tangential vortex ring, with magnitude inversely proportional to the
radius of curvature.}%
\end{center}
\end{figure}

When two vortex lines cross, they will reconnect, as illustrated in Fig.3.
In the final configuration, there appear two cusps, which spring away from
each other at great speed, because of the smallness of the radii of
curvature. A reconnection thus creates two jets, which represents an
efficient way to convert potential energy into kinetic energy in a short
time. The reconnection of magnetic flux lines in the sun's corona is
believed to create jets of solar flares, as shown in Fig.3.

\begin{figure}
[ptb]
\begin{center}
\includegraphics[
height=1.8in,
width=3.4765in
]%
{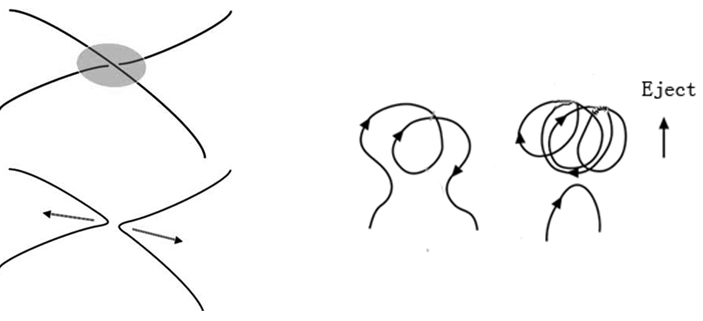}%
\caption{Left panel: Reconnection of vortex lines
creates two cusps that spring away from each other at high velocity,
creating high-velocity jets of flow. Right panel: Jets in magnetic
reconnections in the sun's corona creates solar flares.}%
\end{center}
\end{figure}

Vortex rings can be created in a superfluid with a heat source. They would
expand, intersect and reconnect, and reach a steady-state vortex tangle
called quantum turbulence. Fig.4 \cite{Schwarz88} shows a computer
simulation of the process. The vortex tangle is a new geometrical object,
with a fractal of dimension 1.6 \cite{Kivotides01}. When the heat source is
removed, the vortex tangle will decay. The dynamics of growth and decay can
be described phenomenologically by Vinen's equation \cite{Nemirovskii95}%
\begin{equation}
\frac{\partial \ell }{\partial t}=A\ell ^{3/2}-B\ell ^{2}
\end{equation}%
where $\ell $ is the average vortex-line density $\ell $ (length per unit
volume), and $A$ and $B$ are phenomenological parameters, with $A$ governing
the growth, and $B$ the decay of the vortex tangle.

\begin{figure}
[ptb]
\begin{center}
\includegraphics[
height=1.8in,
width=2.54765in
]%
{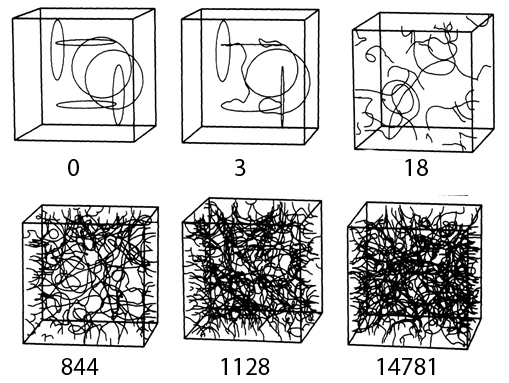}%
\caption{Emergence of quantum turbulence in computer simulation.
Numbers under each picture indicate the number of reconnections. From Ref.%
\protect\cite{Schwarz88}.}%
\end{center}
\end{figure}

In the following, we briefly survey quantum vorticity in diverse physical
systems.

\section{Liquid helium and cold trapped atomic gases}

Quantized vortices in superfluid helium and atomic gases can be described by
the NLSE (\ref{NLSE}), with superfluid density given by $n=\mu _{0}/g$, and
healing length given by%
\begin{equation}
\xi =\frac{\sqrt{g}}{\mu _{0}}
\end{equation}

By shooting alpha particles or electrons into superfluid helium, vortex
rings can be created, which trap the projectiles \cite{Rayfield64}. The
velocity-energy curve of the composite object can be obtained by dragging it
cross the liquid against an electric field. The results fit that of a
quantized vortex ring, as shown in Fig.5.

\begin{figure}
[ptb]
\begin{center}
\includegraphics[
height=2.0in,
width=2.2765in
]%
{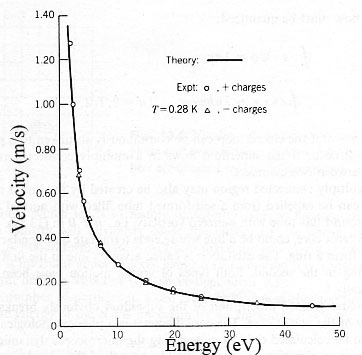}%
\caption{Velocity-energy curve of vortex ring created by ions shot into
liquid helium. The ion (alpha particle or electron) gets trapped in the ring
it created. The theory curve is that for a vortex with one unit of quantized
circulation.}%
\end{center}
\end{figure}

In superfluid helium and cold condensed atomic gases, vortices can be
created by rotating the container. When the angular velocity of the
container exceeds a critical value, the superfluid responds by developing
vortex lines parallel to the axis of rotation \cite{Donelly91} \cite%
{Fetter09}. Experimental results in a trapped atomic gas are shown in Fig.6,
showing the development of a vortex lattice as the angular frequency
increases \cite{Abo02} .

\begin{figure}
[ptb]
\begin{center}
\includegraphics[
height=1.0in,
width=3.4765in
]%
{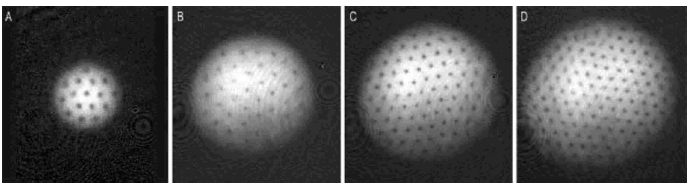}%
\caption{Lattice
of vortices in rotating cold trapped atomic gas, at increasing angular
velocity of rotation. From Ref.\protect\cite{Abo02}.}%
\end{center}
\end{figure}

Quantum turbulence \cite{Paoletti08}, as well as vortex reconnections \cite%
{Paoletti10}, have been studied in superfluid helium. The velocity
distribution in quantum turbulence is found to have a power-law tail $v^{-3}$%
, as shown in Fig.7. This is quite different from the Gaussian distribution
of classical turbulence, and is due to the occurrence of large velocities in
vortex reconnections. Computer simulations using the NLSE have reproduced
this distribution \cite{White10}.

\begin{figure}
[ptb]
\begin{center}
\includegraphics[
height=1.7in,
width=3.4765in
]%
{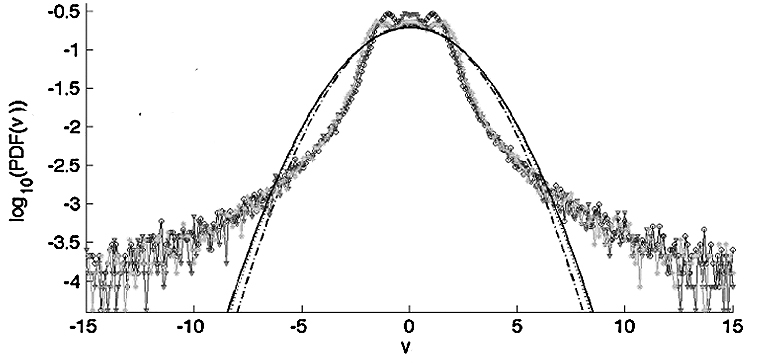}%
\caption{Velocity distribution in quantum turbulence in superfluid helium (data
points), as compared with the Gaussian distribution of classical turbulence
(solid curves). The former has a $v^{-3}$ tail due to the jets of large
velocities accompanying vortex reconnections, which are essential for the
maintenance of quantum turbulence. From Ref. \protect\cite{Paoletti08}.}%
\end{center}
\end{figure}

Vortex lines in liquid helium have been made visible by coating them with
nano-sized metallic dust, as shown in Fig.8 \cite{Lebedev11}. The metallic
dust adheres to the surface of the vortex tube because of the Bernoulli
pressure arising from a decrease of superfluid velocity away from the the
core.

\begin{figure}
[ptb]
\begin{center}
\includegraphics[
height=1.6in,
width=3.4765in
]%
{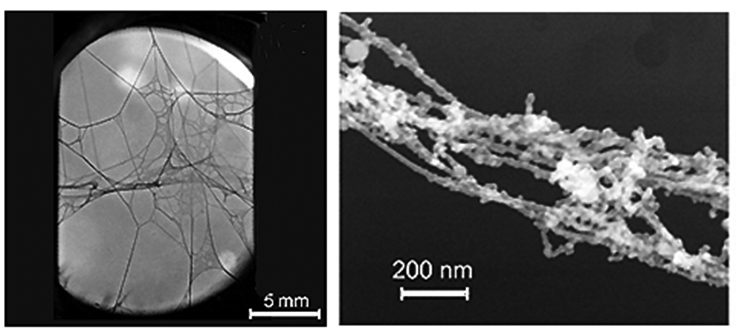}%
\caption{Vortices in superfluid
helium made visible by coating the surface of the vortex with metallic dust
of nano size. The adhesion is due to Bernoulli pressure. From Ref.
\protect\cite{Lebedev11}. }
\end{center}
\end{figure}

The computer simulations in Fig.9 \cite{Berlof02} demonstrates that a Bose
condensate far from equilibrium creates quantum turbulence, and relaxes
through its decay.

\begin{figure}
[ptb]
\begin{center}
\includegraphics[
height=1.8in,
width=3.4765in
]%
{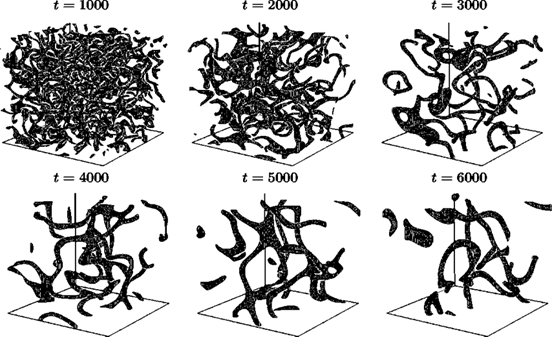}%
\caption{Computer simulation of the evolution of a Bose condensate far from equilibrium,
showing that it equilibrates through the creation and decay of quantum
turbulence. These pictures show the decay sequence. From Ref.\protect\cite%
{Berlof02}.}%
\end{center}
\end{figure}

\section{Superconductivity}

In superconductivity, as described by the GL equation (\ref{GL}), there is
another length besides the healing length $\xi ,$ the penetration depth $d.$
This arises from the conserved current%
\begin{equation}
\mathbf{J}=\frac{\hbar q}{2mi}\left( \phi \nabla \phi ^{\ast }-\phi ^{\ast
}\nabla \phi \right) -\frac{q^{2}}{mc}\phi ^{\ast }\phi \mathbf{A}
\end{equation}%
When this is substituted into Maxwell's equation $\nabla \times \nabla
\times \mathbf{A}=\frac{4\pi }{c}\mathbf{J}$, we obtain, in Coulomb gauge $%
\nabla \cdot \mathbf{A}=0,$%
\begin{equation}
\left( \nabla ^{2}+\frac{4\pi nq^{2}}{mc^{2}}\right) \mathbf{A}=0
\end{equation}%
for uniform $\phi ^{\ast }\phi =n$. The second term in the brackets is the
photon's squared mass inside a superconductor. An external magnetic field
can penetrate a superconductor only to a depth $d$ corresponding to the
inverse mass:%
\begin{equation}
d=\sqrt{\frac{mc^{2}}{4\pi nq^{2}}}
\end{equation}%
This is the Meissner effect, a simple example of the Higgs mechanism, i.e.,
spontaneous breaking of local gauge invariance gives mass to the gauge
particle.

The presence of two characteristic lengths leads two types of
superconductors: If $\xi >d,$ we have type I behavior: the magnetic field is
completely repelled from the interior. If $\xi <d$ we have type II behavior:
the magnetic field penetrates the body in vortex tubes containing quantized
magnetic flux $hc/e$, arranged in a regular lattice know as the Abrikosov
lattice. We can understand these behavior by referring to Fig.10, which
depicts the interface between magnetic field and order parameter. In the
type I case, there is little overlap between magnetic field and order
parameter. If a flux tube were formed, the system can lower the energy by
expelling it outside, and filling in the hole in the order parameter. For
Type II, an overlap between magnetic field and order parameter lowers the
energy, creating a negative surface tension. Thus, flux tubes are formed,
penetrating the superconducting body, and maintained by solenoidal
supercurrents. If the medium is infinite, the flux cannot be expelled, and
finite-energy configurations are flux tube of finite length, terminated by
magnetic monopoles. In this regard, see Fig.11 for a comparison between the
magnetic field due to two monopoles in vacuum and in a superconductor. If
the tube is cut, one does not get free monopoles, but more flux tubes
terminating in $N$ and $S$. In this sense, a superconductor is a medium of
magnetic-monopole confinement.

\begin{figure}
[ptb]
\begin{center}
\includegraphics[
height=2.8in,
width=3.4765in
]%
{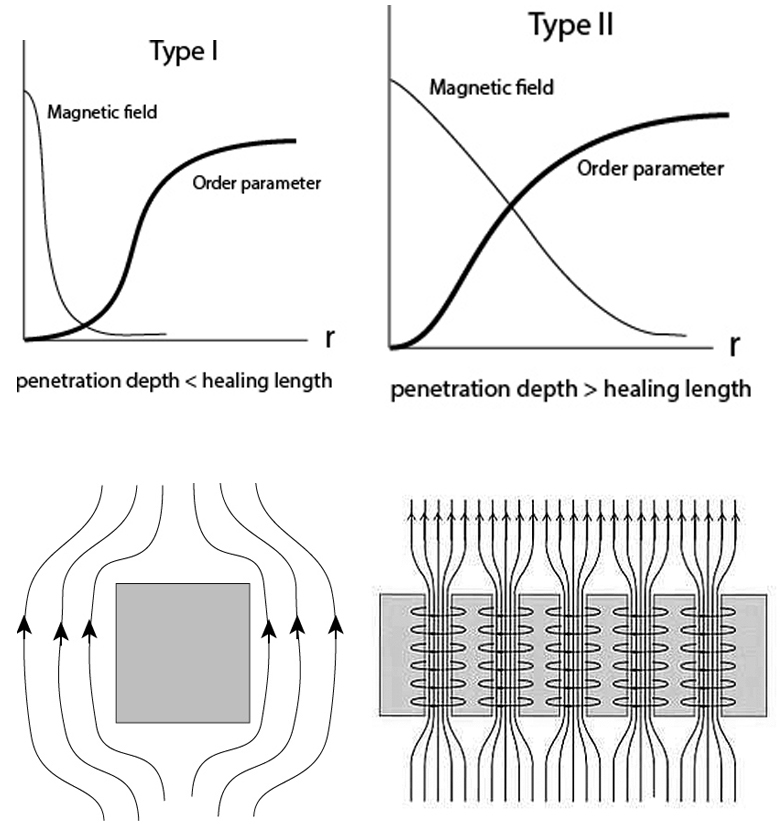}%
\caption{Upper panels show the behavior of the magnetic and order parameter
at an interface, for type I and type II superconductors. lower panels illustrate
the expulsion of magnet flux in type I superconductors, and the formation of a
quantized-flux lattice in type II case.}%
\end{center}
\end{figure}

\begin{figure}
[ptb]
\begin{center}
\includegraphics[
height=1.2in,
width=3.4765in
]%
{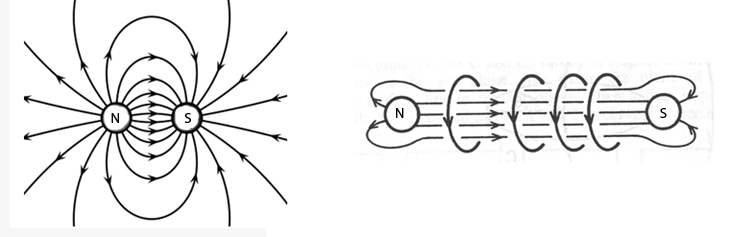}%
\caption{The magnetic field of a
magnetic dipole in empty space (left) and in a superconductor (right). In
the latter the magnetic flux is squeezed into a flux tube by circulation
supercurrents. The situation suggests that a superconductor would exhibit
magnetic monopole confinement.}%
\end{center}
\end{figure}

\section{\textquotedblleft Vorticons\textquotedblright\ in the Higgs field}

In the electroweak sector of the standard model, the local gauge symmetry is
spontaneously broken by of the multi-component Higgs field, which gives mass
to the gauge bosons $W_{\pm }$ and $Z_{0}$, while leaving the photon
massless. We can envisage a "vorticon" \cite{Huang81}, a microscopic vortex
ring in the Higgs field, in the shape of a donut, containing a gauge boson
that is massless, as illustrated in Fig.12. It is a microscopic donut-shaped
waveguide, which could be created when an energetic $Z_{0},$say, tears
through the Higgs field, much as a projectile can create and be trapped by a
vortex ring in liquid helium.

\begin{figure}
[ptb]
\begin{center}
\includegraphics[
height=1.5in,
width=3.4765in
]%
{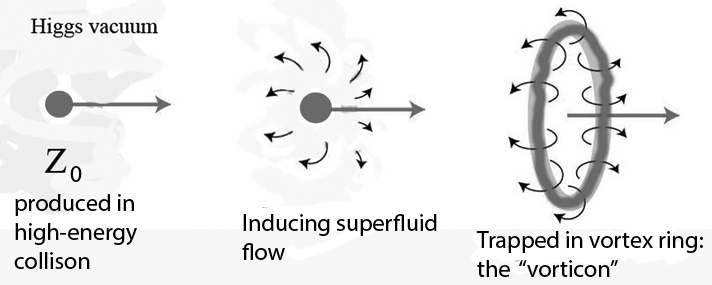}%
\caption{A $%
Z_{0}$ vector boson created in a high-energy collision induces superfluid
flow in the background Higgs field, creating, and is trapped by, of a vortex
ring, resulting in a unstable particle dubbed the \textquotedblleft
vorticon\textquotedblright\ , with a mass of approximately 3 TeV, and a
lifetime of the order of 3$\times $10$^{-25}$s.}%
\end{center}
\end{figure}

The mass of a vorticon can be estimated by constructing normal modes of the $%
Z_{0}$ field inside a torus cut out from the background Higgs field, and
minimizing the energy of the lowest mode \cite{Huang81}. The standard-model
Hamiltonian in the $Z_{0}$ sector is, with $\hbar =c=1$,%
\begin{equation}
H=\int d^{3}x\left[ \frac{1}{2}\left( \mathbf{B}^{2}+\mathbf{E}^{2}\right)
+\left\vert \left( \nabla -iq\mathbf{Z}\right) \phi \right\vert ^{2}+V\left(
\phi \right) \right]
\end{equation}%
Here, $\phi $ is the Higgs field, $\mathbf{Z}$\textbf{\ }the vector
potential in Coulomb gauge $\left( \nabla \cdot \mathbf{Z}=0\right) $, \ and 
$\mathbf{B}=\nabla \times \mathbf{Z},$ $\mathbf{E}=-\partial \mathbf{Z}%
/\partial t.$The Higgs potential is\ given by%
\begin{equation}
V\left( \phi \right) =\frac{\lambda }{2}\left( \phi ^{\ast }\phi
-F_{0}^{2}\right) ^{2}
\end{equation}%
with%
\begin{eqnarray}
F_{0} &=&174\text{ GeV}  \nonumber \\
\sqrt{2\lambda }F_{0} &=&m_{H}=125\text{ GeV}
\end{eqnarray}%
where $m_{H}$ is the Higgs mass. Thus $\lambda =0.256.$ The gauge coupling
constant $q$ and $Z_{0}$ mass are given by%
\begin{eqnarray}
q &=&-\frac{e}{\sin 2\theta _{W}} \\
m_{Z} &=&\frac{gF_{0}}{\sqrt{2}\cos \theta _{W}}
\end{eqnarray}%
where $e$ is given through the fine-structure constant $e^{2}/4\pi \approx
1/137,$ and\ $\theta _{W}$ is the Weinberg angle given through $\sin
^{2}\theta _{W}\approx 1/4$.

There are two types of vorticons: magnetic and electric, with the magnetic
(electric) field pointing along the toroidal direction. As in
electromagnetic wave guides, there is no completely transverse mode. The
masses are found to be 
\begin{eqnarray}
\frac{M_{mag}}{m_{Z}} &=&35.3+6.42\left( \frac{m_{H}}{m_{Z}}\right)
^{2/3}-1.03\left( \frac{m_{H}}{m_{Z}}\right) ^{1/2}  \nonumber \\
\frac{M_{elec}}{m_{Z}} &=&27.7+5.66\left( \frac{m_{H}}{m_{Z}}\right)
^{2/3}+0.504\left( \frac{m_{H}}{m_{Z}}\right) ^{1/2}
\end{eqnarray}%
With the experimental values $m_{Z}=91$ GeV and $m_{H}=125$ GeV, we have%
\begin{eqnarray}
M_{mag} &=&3.47\text{ TeV}  \nonumber \\
M_{elec} &=&2.88\text{ TeV}
\end{eqnarray}%
The size of these vorticons are of order $m_{Z}^{-1}\approx 10^{-12}$ cm.
They are unstable, with lifetimes of the order the $Z_{0}$ lifetime $3\times
10^{-25}$s.

\section{QCD strings}

A phenomenological description of quark confinement in QCD can be modeled
after the magnetic monopole confinement in a superconductor. The difference
is that quarks generate Yang-Mills color-electric fields instead of magnetic
fields, so the QCD vacuum confines electric flux instead of magnetic flux.
(In the nonlinear Yang-Mills gauge theory, there is no electric-magnetic
duality as in Maxwell theory.) A meson consists of a quark and antiquark
pair, belonging respectively to the representations $\mathbf{3}$ and $%
\mathbf{\bar{3}}$ of color SU(3), connected by a flux tube. This can be
represented by the picture in Fig.11., if we substitute $\mathbf{3}$ for $N$%
, $\mathbf{\bar{3}}$ for $S$. The order parameter corresponds to a
condensate of quark-antiquark pairs, and the electric flux tube is
maintained by solenoidal color-magnetic currents. A baryon made of three
quarks can be represented by a flux tube with one quark at one end, at two
at the other. The two-quark system has color representations $\mathbf{3}%
\times \mathbf{3}=$ $\mathbf{\bar{3}}+\mathbf{6},$ and we can use the $%
\mathbf{\bar{3}}$ irreducible combination. Alternatively, the baryon could
be represented by Y-shaped flux tubes each terminated by $\mathbf{3}$.

One can approximate the flux tube by a massless string terminated at both
ends by quarks, with the string rotating about a perpendicular axis, under
tension $T_{0}$ \cite{Johnson79}, as illustrated in the upper panel of
Fig.13. In the limit of small quark masses, one obtains a relation between
between $J$ and $M$:%
\begin{equation}
J=\alpha _{0}+\alpha ^{\prime }M^{2}
\end{equation}%
where $\alpha _{0}$ is a model parameter, and%
\begin{equation}
\alpha ^{\prime }=\frac{1}{2\pi T_{0}}
\end{equation}%
As shown in Fig.13, $J$-$M^{2}$ plots of meson and baryons do yield straight
lines known as Regge trajectories, with a universal slope $\alpha ^{\prime
}\approx 0.9$ GeV$^{-2}$. This corresponds to a string tension of
approximately 16 tons.

\begin{figure}
[ptb]
\begin{center}
\includegraphics[
height=2.1in,
width=3.0765in
]%
{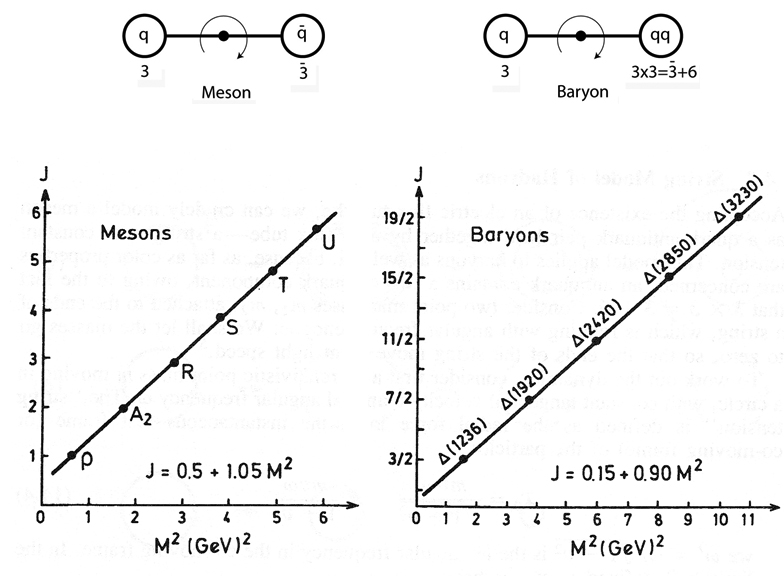}%
\caption{Hadrons
modelled as strings terminated in quarks. The strings are idealization of
vortex tubes containing color-electric flux. Their rotational angular
momentum $J$ and squared mass $M^{2}$ bear a linear relation that agrees
with the observe Regge trajectory of particles, and show the string has a
tension of about 16 tons.}%
\end{center}
\end{figure}

It would be desirable to set up an analog of the GL equation for QCD, but
the situation here is more complicated than superconductivity, because of
the intrinsic non-linearity of Yang-Mills gauge fields. Besides quark
confinement, there is chiral symmetry breaking. The condensate could involve
topological object such as monopoles. We refer the reader to \cite{Huang92}, 
\cite{Gusynin92}, \cite{Xiong13}, for works related to that goal.

\section{Cosmology{}}

The Higgs field of particle theory has found experiment support with the
discovery of the Higgs boson \cite{ATLAS12}, \cite{CMS12}. Grand-unified or
supersymmetric theories call for more Higgs-like fields. Any long-range
order, such as the phase coherence of these fields, will persist to infinite
distances in spatial dimension greater than two.  In two or fewer
dimensions, the long-range order will be destroyed by the long-wavelength
fluctuations of the Goldstone modes \cite{Huang87}. The Higgs field, and
other possible Higgs-like fields, thus make the whole universe a
superposition of different types of superfluid, (like a mixture of $^{4}$He
and $^{3}$He below 10$^{-3}$K.) We discuss the manifestations of
superfluidity in terms of a generic complex scalar field, in three different
epochs of the universe: the big-bang era, the CMB (cosmic microwave
background) era, and the present.

The scalar field presumably emerges during the big bang, but how it does
this depends on model. A quantum field needs a high-momentum cutoff $\Lambda
,$ which should be infinite at the big bang. If the potential emerges from
zero at that moment, it must be asymptotically free. This rules out all
polynomial forms, and admits only the Halpern-Huang potential \cite%
{Halpern96}, with exponential behavior for large fields. A big-bang model
has been constructed based on this potential \cite{Huang12a} \cite{Huang12b}%
, with uniform Robertson-Walker metric whose scale is tied to the cutoff: $%
a=\Lambda ^{-1}$. Solving Einstein equation numerically, as an initial-value
problem, shows that the length scale expands like 
\begin{equation}
a\left( t\right) \sim \exp \left( c_{0}t^{1-p}\right)
\end{equation}%
where $c_{0}$ and $p<1$ depend on model parameters and initial data. The
result is equivalent to having a cosmological constant that decays in time
according to a power law. If this behavior persists to later times, it could
explain dark energy without "fine-tuning". Phenomenological studies so based
are referred to as "intermediate inflation" \cite{Barrow95}.

During the big bang, the scalar field would emerge far from equilibrium,
and, like the Bose gas illustrated in Fig.9, equilibrate by going through a
period of quantum turbulence. The vortex reconnections that maintain the
vortex tangle produce jets of energy that can create matter, like the solar
flares created by magnetic reconnections illustrated in Fig.3. This provides
a new scenario for the inflation era, in which all matter in the present
universe were created in quantum turbulence.

A detailed analysis of the turbulent era based on the coupled
Einstein-scalar equations, unfortunately, faces the formidable problem of a
non-uniform metric. We can bypass this difficulty with a phenomenological
approach using Vinen's equation, which deals with a uniform distribution of
vortex lines. It is then possible to construct a scenario for the growth and
decay of a vortex tangle, as illustrated in Fig.14. Parameters can be so
chosen that the lifetime of the vortex tangle is of the order of 10$^{-26}$%
s, during which time the universe expanded by a factor of 10$^{27}$, and a
total amount of matter was created equal to what is presently observed,
roughly 10$^{22}$ suns. This presents a new picture of inflation \cite%
{Huang12b}.

\begin{figure}
[ptb]
\begin{center}
\includegraphics[
height=2.4in,
width=3.4765in
]%
{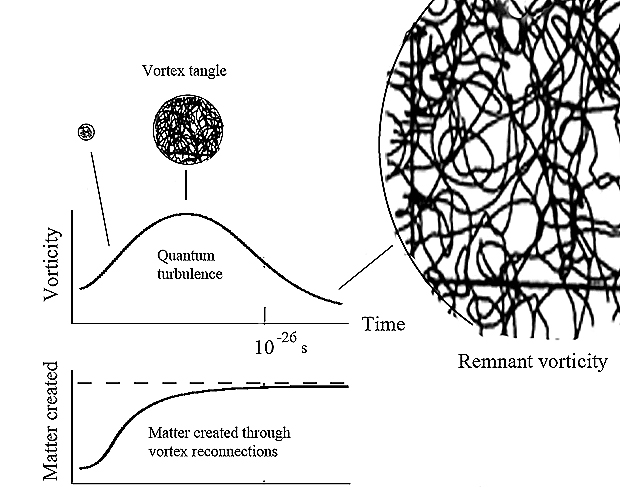}%
\caption{The inflation era:
creation of matter all matter the vortex reconnections in quantum
turbulence. Ref.\protect\cite{Huang12b}.}%
\end{center}
\end{figure}

The big-bang era, including inflation and the creation of all matter through
quantum turbulence, lasted about 10$^{-26}$s. After this, nonuniformities in
the universe become important, and the Robertson-Walker metric ceases to be
valid.\ The Halpern-Huang potential, which is a high-cutoff approximation,
may need corrections. Also, while there was only one scale in the big-bang
ear, the Planck scale of 10$^{22}$ GeV, a new scale appears with the
emergence of matter, the QCD scale of 1 GeV. Vortices created during the big
bang era, including those in the vortex tangle, have core sizes of the order
of the Planck length of 10$^{-33}$cm, and they will co-expand with the
universe. Those created after the emergence of the matter, however, will
have much smaller fixed core sizes that do not expand with the universe.

After the big bang era, our model will be replaced by a cosmic perturbation
theory, which must include the scalar field, and this will govern the
creation of the CMB. The widely used $\Lambda CDM$ model would be modified
with inclusion of a vacuum complex scalar field, which supplies a
cosmological-constant $\left( \Lambda \right) $ via its energy-momentum
tenor, and generates cold dark matter $\left( CDM\right) $ through density
variations of the scalar field. A complete reformulation is yet to be done,
but we expect that the results of the usual cosmic perturbation theories
will be largely preserved. The only new element would be quantum vorticity,
which will make contributions to the tensor mode, in competition with
gravitational waves.

In the present universe, galaxies will attract superfluid to form halos of
denser-than-vacuum superfluid around it, and these will be perceived as dark
matter, through gravitational lensing \cite{Huang14}. When galaxy clusters
collide, their haloes undergo distortions according to superfluid
hydrodynamics, as observed in the "bullet cluster" \cite{Crowe06}. Computer
simulations based on the NLKG may be found in \cite{Huang14}. The literature
contains studies of cosmic vorticity under the name "cosmic strings" \cite%
{Gradwohl90} \cite{Villenkin94}.

\begin{figure}
[ptb]
\begin{center}
\includegraphics[
height=3.2in,
width=3.4765in
]%
{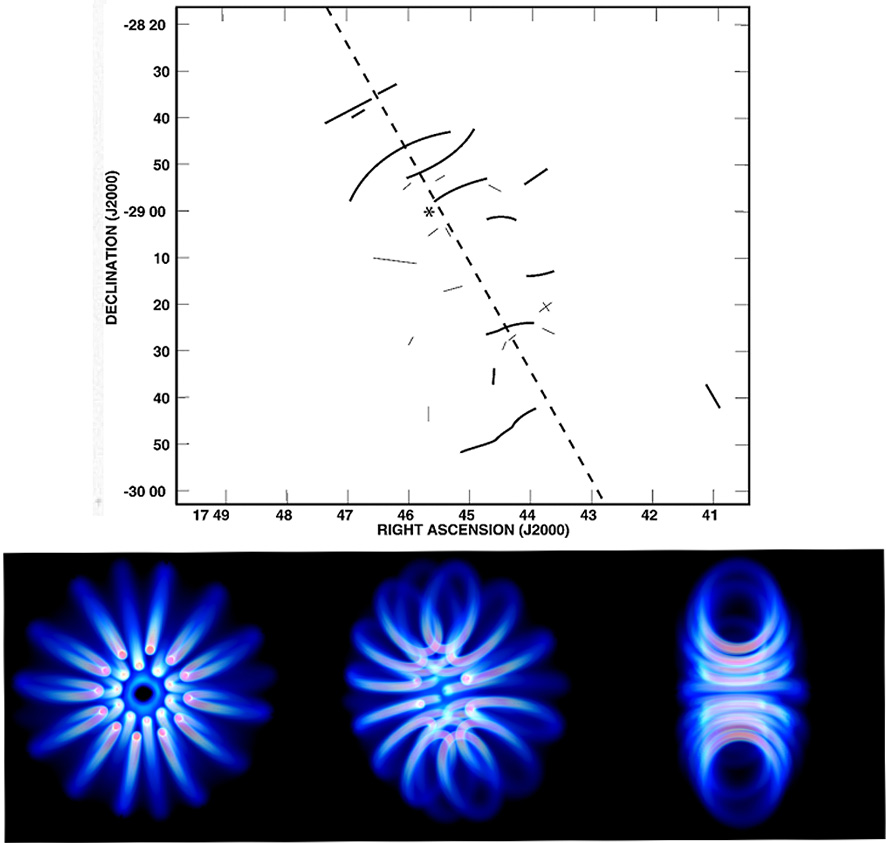}%
\caption{Upper panel: The "non-themal
filaments" observed near the center of the Milky Way, Ref.\protect\cite%
{LaRosa04}, could be vortex lines surrounding black holes. Lower panel: 3D
simulations of a vortex-ring assembly arising from a rotating body at the
center, based on the NLKG, in various perspectives, Ref.\protect\cite%
{Xiong14}.}%
\end{center}
\end{figure}

A fast-rotating body such as a black hole will drag the surrounding
superfluid into rotation through the creation of vortices. The so-called
"non-thermal" filaments \cite{LaRosa04} observed near the center of the
Milky Way could be vortex lines, as shown in the upper panel of Fig.15. The
lower panels shows 3D simulations based on the NLKG \cite{Xiong14}. \ 

The core sizes of remnant vortices from quantum turbulence in the big bang
era, originally of order of the Planck size of 10$^{-33}$cm, will co-expand
with the universe, and in the 13.7 billion years since, they could reach
sizes of the order of 10$^{7}$ light years. Since matter was created in the
scalar field, these cores are devoid of matter, and show up as voids in the
galactic distribution. The so-called "stick man" configuration \cite%
{deLapparent86} is shown in Fig.16, with a simulation by superposition of
vortex tubes.

\begin{figure}
[ptb]
\begin{center}
\includegraphics[
height=1.5in,
width=3.4765in
]%
{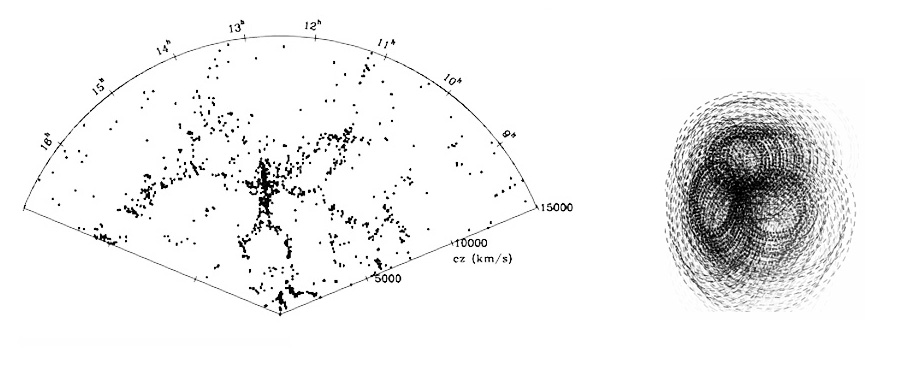}%
\caption{Left: Voids in galaxy
distribution in the "stick man" configuration, Ref. \protect\cite%
{deLapparent86}. Right: Simulation by the superposition of three primordial
vortex tubes, which have co-expanded with the universe to gigantic sizes.
The inside of the tubes are devoid of galaxies, which cling to the outside
dues to hydrodynamic pressure, like the metallic dust clinging to vortex
tubes in superfluid helium shown in Fig.8.}%
\end{center}
\end{figure}

\section{Galaxy formation}

The CMB was formed about 10$^{5}$ yrs after the big bang. Between that and
the present, there was a long period of galactic formation, in which quantum
vorticity may play an important role. A speculation of Lathrop \cite{Lathrop}
is that a galaxy can form from a large vortex ring with accretion of dust,
which gravitates to form a central mass, squeezing the vortex ring into a
shape with spiral arms, as illustrated in Fig.17. Can one find any hint of
this mechanism in currently-observed galactic properties?

\begin{figure}
[ptb]
\begin{center}
\includegraphics[
height=1.3in,
width=3.4765in
]%
{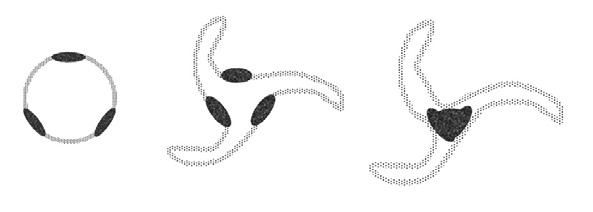}%
\caption{Lathrop's suggestion for galaxy formation:
dust particles accrete onto a vortex ring, gravitate and clump, squeezing
the ring into a spiral shape with a central mass, which would become a
black hole.}%
\end{center}
\end{figure}

The central mass would correspond to the black hole observed at the center
of all galaxies, whose mass $M$ bears power-law relations to other galactic
properties \cite{Nicholas12}:%
\begin{equation}
M\sim X^{\beta }
\end{equation}%
as indicated in the following:%
\begin{equation}
\begin{array}{ccc}
& X & \beta \\ 
\text{Stellar mass} & m & 1.05\pm 0.11 \\ 
\text{Luminosity} & L & 1.11\pm 0.13 \\ 
\text{Stellar velocity} & v & 5.57\pm 0.33%
\end{array}%
\end{equation}

Assume that the dust particles initially accrete onto the vortex ring
uniformly, and then clump up under gravitation. Assume further that a fixed
fraction forms the central mass, which\ becomes a black hole and goes dark,
while the rest remains luminous. This would mean $m\varpropto $ $M$, and $%
L\varpropto m$, which are consistent with the relevant exponents being unity.

The initial vortex ring may be generated by self-avoiding random walk (SAW)
. (The fact that it is a close ring matters little for the arguments here.)
Thus, the dust particles, which initially adhere uniformly to the vortex
ring, form a SAW sequence, with the relation \thinspace $N\sim R^{5/3},$
where $R$ is the spatial extension, and $N$ the number of steps. {}In our
case, $R$ corresponds to the size of the galaxy, and $N$ is proportional to $%
M+m$, hence to $M$ . This means that $M$ scales like $R^{5/3}$. Now assume
that the total angular momentum $J$, which is a constant of the motion,
scales as the galactic volume:

\begin{eqnarray}
M &\sim &R^{5/3}  \nonumber \\
J &\sim &R^{3}
\end{eqnarray}%
Defining the stellar velocity $v$ through $J=Rmv\varpropto $ $RMv$, we obtain%
\begin{equation}
M\sim v^{5}
\end{equation}%
which is not inconsistent with observations. As an interesting note, the SAW
exponent 5/3 is the same as the Kolmogorov exponent in classical turbulence,
and the Flory exponent for polymers \cite{Huang05}.

\end{document}